\documentclass[fleqn,usenatbib]{mnras}

\usepackage[T1]{fontenc}

\DeclareRobustCommand{\VAN}[3]{#2}
\let\VANthebibliography\thebibliography
\def\thebibliography{\DeclareRobustCommand{\VAN}[3]{##3}\VANthebibliography}

\usepackage{graphicx}	
\usepackage{amsmath}	
\usepackage{newtxtext,newtxmath} 

\usepackage[colorinlistoftodos]{todonotes}
\usepackage[font=small,skip=-5pt]{caption}

\usepackage{soul} 
\usepackage{xargs} 

\newcommand{\amm}{NH$_{3}(1_{0}-0_{0})$}

\title[Ammonia at $z=2.6$]{Ammonia in the interstellar medium of a starbursting disc at $\mathbf{z=2.6}$  }

\author[M. J. Doherty et al.]{
M. J. Doherty,$^{1}$\thanks{E-mail: m.doherty2@herts.ac.uk}
J. E. Geach,$^{1}$
R. J. Ivison,$^{2}$ 
K. M. Menten,$^{3}$ 
A. M. Jacob,$^{4}$
J.~Forbrich,$^{1}$ \&
S. Dye$^{5}$
\\
$^{1}$Department of Physics, Astronomy \& Mathematics, University of Hertfordshire, College Lane, Hatfield, AL10~9AB\\
$^{2}$European Southern Observatory, Karl-Schwarzschild-Straße 2, D-85748 Garching, Germany\\
$^{3}$Max Planck Institute for Radio Astronomy, Auf dem H{\" u}gel 69, 53121 Bonn, Germany\\
$^{4}$William H. Miller III Department of Physics \& Astronomy, Johns Hopkins University, Baltimore, MD 21218, USA\\
$^{5}$School of Physics and Astronomy, University of Nottingham, University Park, Nottingham NG7~2RD, UK
}

\pubyear{2022}

\begin{document}
\label{firstpage}

\pagerange{\pageref{firstpage}--\pageref{lastpage}}
\maketitle

\begin{abstract}
We report the detection of the ground state rotational emission of ammonia, ortho-NH$_3$\,$(J_K=1_0\rightarrow0_0)$ in a gravitationally lensed, intrinsically hyperluminous, star-bursting galaxy at $z=2.6$. The integrated line profile is consistent with other molecular and atomic emission lines which have resolved kinematics well-modelled by a 5\,kpc-diameter rotating disc. This implies that the gas responsible for NH$_3$ emission is broadly tracing the global molecular reservoir, but likely distributed in pockets of high density ($n\gtrsim5\times10^4$\,cm$^{-3}$). With a luminosity of $2.8\times10^{6}$\,$L_\odot$, the NH$_3$ emission represents $2.5\times10^{-7}$ of the total infrared luminosity of the galaxy, comparable to the ratio observed in the Kleinmann-Low nebula in Orion and consistent with sites of massive star formation in the Milky Way. If $L_{\rm NH_3}/L_{\rm IR}$ serves as a proxy for the `mode' of star formation, this hints that the nature of star formation in extreme starbursts in the early Universe is similar to that of Galactic star-forming regions, with a large fraction of the cold interstellar medium in this state, plausibly driven by a storm of violent disc instabilities in the gas-dominated disc. This supports the `full of Orions' picture of star formation in the most extreme galaxies seen close to the peak epoch of stellar mass assembly.
\end{abstract}




\section{Introduction}

The progenitors of the most massive galaxies today are most likely the population of intense star-bursting galaxies seen at $z\gtrsim2$ \citep{smail_deep_1997,hughes_high-redshift_1998}. These starbursts have large gas reservoirs \citep{bothwell_survey_2013} representing a significant fraction of baryonic mass \citep{wiklind_evolution_2019} fuelling very high rates of star formation, possibly up to three orders of magnitude greater than the Milky Way \citep{chapman_evidence_2004,Barger_2014}.

Generally, these gas- and dust-rich systems are obscured in the optical, but radiate strongly in the submillimetre and millimetre through their thermal dust emission. Indeed, the emission from key molecular and atomic tracers of the cool and cold dense interstellar medium (ISM) responsible for fuelling star formation, and its immediate environment is also observed at these wavelengths.

Two key factors have improved our understanding of the nature of these distant, dusty, prodigiously star-forming galaxies over the past decade. One is the identification of large samples of strongly gravitationally lensed systems \citep[e.g.\,][]{danielson_13co_2013,spilker_rest-frame_2014,rybak_alma_2015,geach_red_2015,rybak_full_2020}. Lensing amplifies flux, revealing emission features otherwise too faint to detect, and magnifies images of galaxies to provide access to spatial scales not achievable by any other means \citep{rybak_alma_2015,geach_magnified_2018,rybak_full_2020}. The second is the advent of sensitive, wide bandwidth interferometry across the submillimetre--millimetre using large interferometric arrays, in particular the Atacama Large Millimetre/submillimetre Array (ALMA). The follow-up of lensed galaxies with ALMA has accelerated the establishment of  a clearer, albeit still incomplete, picture of the nature of galaxies undergoing intense starbursts in the early Universe.

\begin{figure*}
	\centering
	\includegraphics[width=0.8\textwidth]{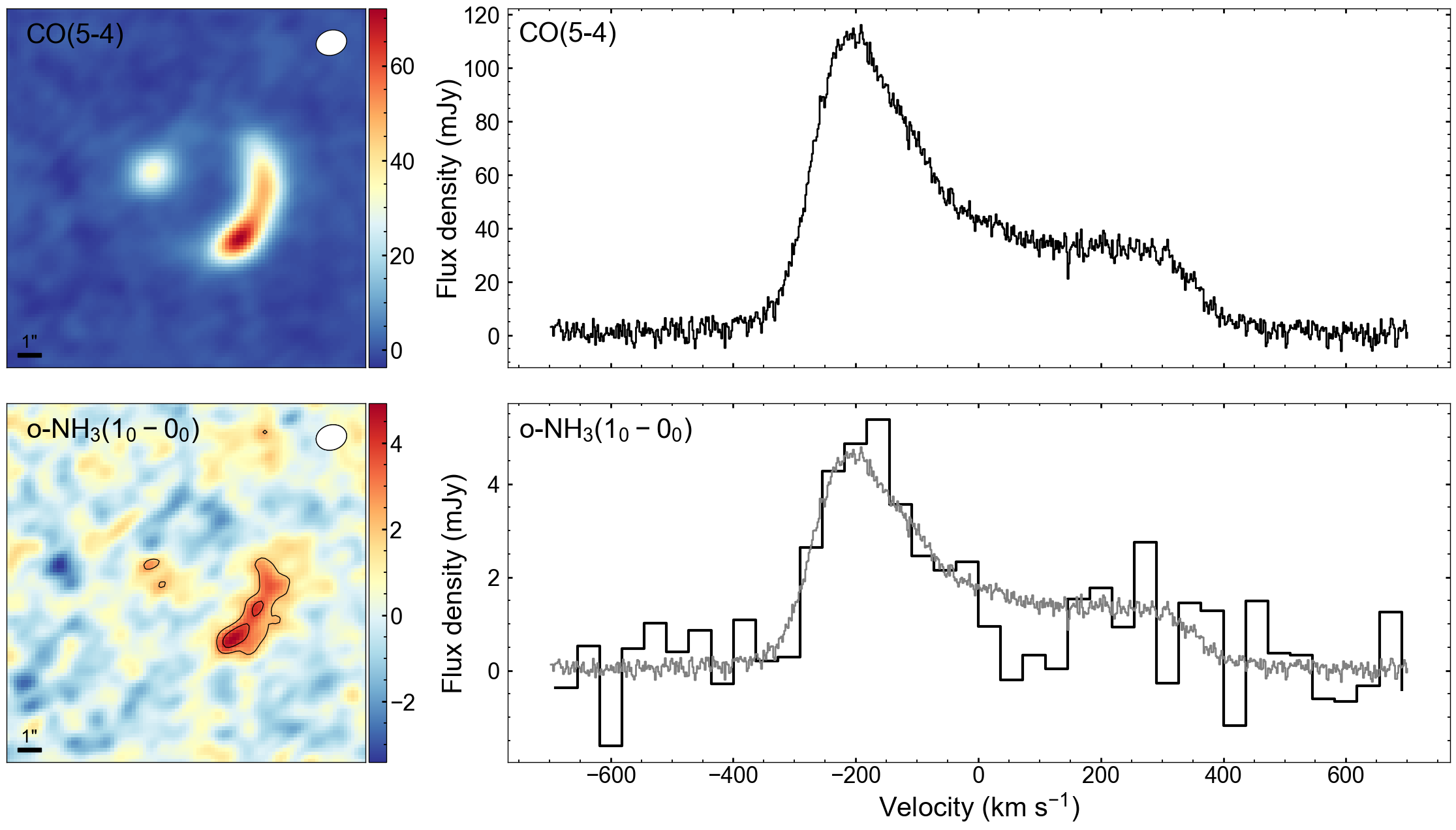}
    \caption{Maps and spectra for the NH$_{3}$ and CO(5--4) emission in 9io9. (left) Maps show the continuum-subtracted image plane emission averaged over the line with contours shown for the fainter NH$_3$ starting at 3$\sigma$. Synthesised beams are indicated top right. (right) Spectra show the source-integrated emission, with the much brighter CO line acting as a reference profile that can be directly compared with the weaker NH$_3$, where we overlay a scaled version of the CO(5--4) line to demonstrate the strong similarity between the line profiles, despite NH$_3$ generally tracing much denser gas (see Section 4 for a discussion). We bin the NH$_3$ spectrum to $\sim$50\,km\,s$^{-1}$ channels for visualisation purposes. The `dip' in NH$_{3}$ emission at $\sim$100~km~s$^{-1}$ is not a significant feature.}
    \label{fig:NH3}
\end{figure*}
What astrophysics is responsible for the existence of extreme starbursts at high-{\it z}? Compared to the merger-dominated ultra luminous infrared galaxies (ULIRGs) in the local universe with intense circumnuclear star formation, high-{\it z} starbursts of equivalent luminosity appear to be sustaining star formation across much larger scales \cite{rujopakarn_morphology_2011}. 

Mergers undoubtedly play a role in the triggering high-{\it z} starbursts  \citep{tacconi_high_2010,engel_most_2010}, but there is some observational evidence that some of the most intensely star-forming galaxies are simply consistent with large gas-dominated rotationally supported discs \citep{hodge_kiloparsec-scale_2016,jimenez-andrade_molecular_2018,geach_magnified_2018,gullberg_dust_2018}. In these systems a significant fraction of the ISM appears to have been driven to high density \citep{oteo_high_2017,geach_magnified_2018,doherty_n_2020}, possibly via violent disc instabilities \citep[VDIs, e.g.][]{toomre_gravitational_1964, dekel_formation_2009,dekel_cold_2009}, and in this case the extreme star formation rates measured can naturally be explained by the sheer quantity of gas available for active participation in star formation \citep{geach_molecular_2012,papadopoulos_molecular_2012}. We can learn more about the actual conditions of the star-forming ISM, and the conditions of star formation in general by exploiting gravitational lensing to study the astrochemistry of these systems \citep{danielson_properties_2011, spilker_rest-frame_2014, zhang_stellar_2018,dye_high-resolution_2022}. If the density distribution of the cold molecular reservoir is important for driving globally high star formation rates, then observations of species that trace the densest environments are required, particularly heavy rotor molecules \citep{oteo_high_2017,bethermin_dense-gas_2018}.

In this work we report the detection of the ground state emission line of ammonia (NH$_3$) in a (now well-studied) strongly lensed starburst galaxy at $z=2.6$. A tracer of the dense molecular ISM and intimately linked to the sites of star formation, NH$_3$ was the first polyatomic molecule detected in the ISM \citep{cheung_detection_1968} and is amongst the most studied species in the local universe, primarily through its radio inversion lines \citep{ott_ammonia_2011,schmidt_herschel_2016,feher_ammonia_2022}. The ground state NH$_{3}$\,($J_K=1_0\rightarrow0_0$) ortho line emits at 572.498\,GHz in the rest frame, and therefore any ground-based studies of this feature at $z=0$ are hampered by the near-zero transmission of the atmosphere at this frequency. In Section 2 we present the observations and data reduction, in Section 3 we present our analysis and results, and in Sections 4 and 5 we provide our interpretation and conclusions. Throughout we assume a `Planck 2015' cosmology where $H_{0} = 68~\text{km}~\text{s}^{-1}\text{Mpc}^{-1}$ and $\Omega_{\text{m}} = 0.31$ \citep{planck_collaboration_planck_2016}.

\section{Observations and data reduction}

9io9 (J2000, $02^{\rm h}09^{\rm m}41\fs3$, $00\degr15\arcmin58\farcs5$, $z=2.5543$) was observed with the ALMA 12\,m array during project 2019.1.01365.S. The C43-3 configuration was used, employing 48 antennas with baseline separations of 15--784\,m. We executed a spectral scan in Band 4 across $\nu_{\rm obs}=152.5\text{--}162.9$\,GHz with a total on-source time of approximately 195\,minutes over five executions. The precipitable water vapour column was 1.9--3.5\,mm and the average system temperature was $T_{\rm sys}=69\text{--}91$\,K over the five executions. Atmosphere, bandpass, phase and pointing calibrators included the sources J0423$-$0120 and J0217+0144. We use the pipeline-restored calibrated measurement set for imaging. We image and CLEANed the data using {\sc casa} (v.5.1.0-74.el7) {\tt tclean} with multiscale cleaning at scales of $0\arcsec$, $0.5\arcsec$, and $1.25\arcsec$. First we produce dirty cubes to establish the r.m.s. (1$\sigma$) noise per channel, and then cleaned down to a stopping threshold of 3$\sigma$. With natural weighting, and setting a common beam to the whole datacube, the synthesized beam has a full width at half maximum of $1.3\arcsec \times 1.0\arcsec$ (position angle $72\degr$). The r.m.s. noise per 10\,MHz (20\,km\,s$^{-1}$) channel is 0.3\,mJy\,beam$^{-1}$.   

We adopt the same lens model as \cite{geach_magnified_2018}. Briefly, the
lens model includes the gravitational potential of both the primary lensing
galaxy ($z\approx0.2$) and its smaller northern companion (assumed to be at
the same redshift). The model uses the semi-linear inversion method of
\cite{warren_semilinear_2003} to reconstruct the source plane surface brightness that best matches the
observed Einstein ring for a given lens model. This process is iterated,
varying the lens model and creating a source reconstruction at each
iteration, until the global best fit lens model is found \citep{geach_magnified_2018}. The best fit model is used to produce source-plane
cubes. In the following, we use the source-plane cube to extract the integrated spectrum, accounting for magnification of the sources of line emission. However, due to the relatively coarse resolution and signal to noise of the data, we present maps of 9io9 in the image plane.

\section{Results}

We use {\it Splatalogue} \citep{remijan_splatalogue_2007} to identify emission lines in the total spectrum. CO $J = 5\rightarrow4$ at $\nu_{\rm obs}=162.133$\,GHz is detected at high significance as expected, and exhibits the characteristic double horned profile as other molecular and atomic lines \citep{geach_red_2015,geach_magnified_2018,harrington_red_2019,doherty_n_2020}, and well-modelled in the source-plane reconstruction by a rotating disc \citep{geach_magnified_2018}. We also detect a fainter, but significant, emission feature at  $\nu_{\rm obs}=161.072$\,GHz that is consistent with the redshifted ground state ortho-NH$_3$\,$J_K=1_0 \rightarrow 0_0$ rotational line at $\nu_{\rm rest}=572.498$\,GHz \citep[][hereafter we  refer to the line as NH$_3$]{cazzoli_hyperfine_2009}. Image plane integrated spectra of the CO $J = 5\rightarrow4$ (hereafter CO(5--4)) and NH$_3$ lines and maps are presented in Figure~\ref{fig:NH3}. To our knowledge, the previous highest redshift detection of this transition of ammonia was in absorption in a spiral galaxy at $z=0.89$, where the galaxy is acting as a lens, magnifying the strong (sub)millimeter continuum emission  of the famous background quasar PKS 1830$-$211 at $z=2.51$ \citep{menten_submillimeter_2008, muller2014}. 

To measure the line properties, we subtract continuum emission on a pixel-by-pixel basis, using a simple linear fit to the spectrum in line-free regions around NH$_3$. We then model the integrated line emission using an empirical template based on the high SNR CO(5--4) emission line. 
By simply shifting the position of the CO(5--4) emission in frequency space and scaling its amplitude, we can minimise the \smash{$\chi^{2}$} difference between the scaled CO(5--4) and the NH$_3$ line. Figure~\ref{fig:NH3} shows how the scaled CO(5--4) line provides an excellent fit to the NH$_3$ emission. We discuss the implications of this later. The total integrated line flux is evaluated by summing the flux within an aperture defined by the 3$\sigma$ contour of the averaged band 4 data cube. We measure $\mu L_{\rm NH_3}=(3.3\pm 0.2)\times10^7 L_\odot$, where $\mu$ is the lensing magnification. To estimate uncertainties in this procedure, we simply add Gaussian noise to each channel, with a $\sigma$ determined from the r.m.s. in off-source regions of the data cube and then repeat the fit 1000 times. Applying this same procedure to the source plane reconstructions, we obtain a source plane line luminosity of $L_{\rm NH_3}=(2.8\pm 0.2)\times10^6 L_\odot$. If instead of using the scaled CO(5--4) as a model of the emission, we just integrate over the range $\Delta V = \pm 500$~km~s$^{-1}$, we obtain a source plane luminosity of    $L_{\rm NH_3}=(3.1\pm 0.3)\times10^6 L_\odot$.  

\section{Interpretation}

As can be seen from Figure~\ref{fig:NH3}, the scaled CO(5--4) emission is an excellent description of the NH$_3$ line emission. In turn, the line profile of the integrated, projected CO emission is well modelled by a nearly edge-on rotating disc (when modelled in the source plane and projected into the image plane) and this profile is shared by the majority of detected lines within this system covering a wide range of conditions, from the relatively low density molecular reservoir traced by C~{\sc i}\,(1--0) to the warmer, dense, ionised gas traced by N$^+$ \citep{su_redshift_2017,geach_magnified_2018,harrington_red_2019,doherty_n_2020}. The striking similarity in the observed line profiles imply that the observed NH$_{3}$ emission is broadly co-located with the CO-emitting gas, likely emanating from discrete sites of on-going star formation scattered throughout the gas-rich disc.  

\cite{ho_interstellar_1983} note that NH$_3$ is a rather ubiquitous molecule, tracing a broad range of interstellar environments containing molecular gas. However, for the rotational transitions in the millimetre, NH$_3$ is expected to trace dense gas. In the optically thin limit, the critical density of NH$_3$ is $n_{\rm crit}\gtrsim10^7$\,cm$^3$ for kinetic temperatures $T_k<100$\,K \citep{shirley_critical_2015}. In realistic scenarios, the NH$_3$ emission will be optically thick, and therefore subject to radiative trapping \citep{draine_physics_2011}. This serves to lower the effective critical density, but even so, the effective densities for optically thick NH$_3$ emission are still probing dense gas, with $n_{\rm eff}\gtrsim5\times10^4$\,cm$^3$ for $T_k<100$\,K and assuming a column density commensurate with dense gas clumps and cores in the Milky Way ($\log_{\rm 10}(N_{\rm ref}/{\rm cm^{-2}})=14.3$, \cite{shirley_critical_2015}). However, the effective density will further scale down with increasing column density as $N_{\rm ref}/N$. Another caveat is the presence of significant far-infrared background fields, which will be dominated by the ambient radiation field of the galaxy itself due to dust emission, with the most intense emission likely co-located with the dense star-forming gas. This background could lead to significant radiative pumping of NH$_{3}$ molecule and therefore a non-collisional route to rotational emission; indeed \cite{schmidt_herschel_2016} discuss the potential role of pumping of the NH$_3$ rotational ground state emission as a solution to the  discrepancy between the abundances derived via the NH$_{3}$ ground state and its radio inversion lines in some local systems. While radiative pumping would further serve to lower the effective density of the gas responsible for NH$_3$ emission, we can be reasonably confident that the observed NH$_3$ is tracing some of the densest molecular gas in 9io9, and therefore the actual sites of star formation. 

Can we relate the properties of 9io9 to local star formation? \cite{doherty_n_2020} show that the average electron density ($n_{\rm e}\approx 300$\,cm$^{-3}$) associated with warm ionised gas as traced by N$^+$ fine-structure emission is consistent with the typical density of Galactic star-forming regions. The conclusion is that the conditions are not `extreme' compared to sites of active star formation in the Milky Way, but clearly a larger fraction of the ISM is participating in star formation in 9io9 and galaxies like it compared to the Milky Way. What of the efficiency, or `mode' of star formation? A crude approach is to compare proxies for the star formation rate and dense gas that is fuelling it; more efficient star formation is characterised by a higher rate per unit dense gas mass, with a theoretical upper limit set by the Eddington limit \citep{Murray_2005}. With the integrated infrared luminosity as a proxy for the total star formation rate (for galaxies dominated by dust) and NH$_3$ as a  tracer of the dense molecular gas actively participating in star formation, we can use $L_{\rm IR}/L_{\rm NH_3}$ as an empirical tracer of the star formation efficiency. In 9io9 we measure $ L_{\rm NH_3} = 2.8 \times 10^{6} L_{\odot}$ and luminosity of $L_{\rm IR} = 1.1 \times 10^{13} L_{\odot}$, yielding $L_{\rm NH_3}/L_{\rm IR}\approx 3\times 10^{-7}$.

There are relatively few regions where we have robust \amm{} and integrated infrared luminosities. One such region, the Kleinmann–Low nebula in Orion (Orion-KL)  -- a dense, hot molecular cloud core  
close to the Trapezium cluster, which excites the Orion Nebula  -- is frequently used as a local benchmark in many studies, not the least because of its proximity at $\sim$400~pc. With $L_{\rm IR} \sim 8 \times 10^{4}~L_{\odot}$ \citep{gezari_mid-infrared_1998} and  $L_{\rm NH_3} \approx 0.01L_{\odot}$ \citep{olofsson_spectral_2007, Persson2007}, Orion-KL has $L_{\rm NH_3}/L_{\rm IR}\approx 1.3\times 10^{-7}$, within a factor of a few of the `global' 9io9 ratio, despite eight orders of magnitude separating the infrared luminosities.

\begin{table}
\begin{center}
\caption{A comparison of the properties of 9io9 and Galactic sources where NH$_3$ is detected in emission.}
\begin{tabular}{cccc}
\hline
Source & $L_{\rm IR}$ & $L_{\rm NH_{3}}$ &$L_{\rm NH_{3}} / L_{\rm IR}$\\
& $L_{\sun}$ & $L_{\sun}$ & $\times10^{-7}$\\
\hline
9io9 (this work) & $1.1 \times 10^{13}$ & $2.8 \times 10^{6}$ & $2.5$\\
Orion-KL & $8 \times 10^{4}$ & $0.01$ & $1.3$ \\
W31 C & $\sim$$10^{6}$ & $>$0.042 & $>$$0.42$ \\
W49 N & $\sim$$10^{7}$ & $>$0.048 & $>$$0.48$\\
\hline
\label{tab}
\end{tabular}
\end{center}
\end{table}

\begin{figure}
    \centering
    \includegraphics[width=0.425\textwidth]{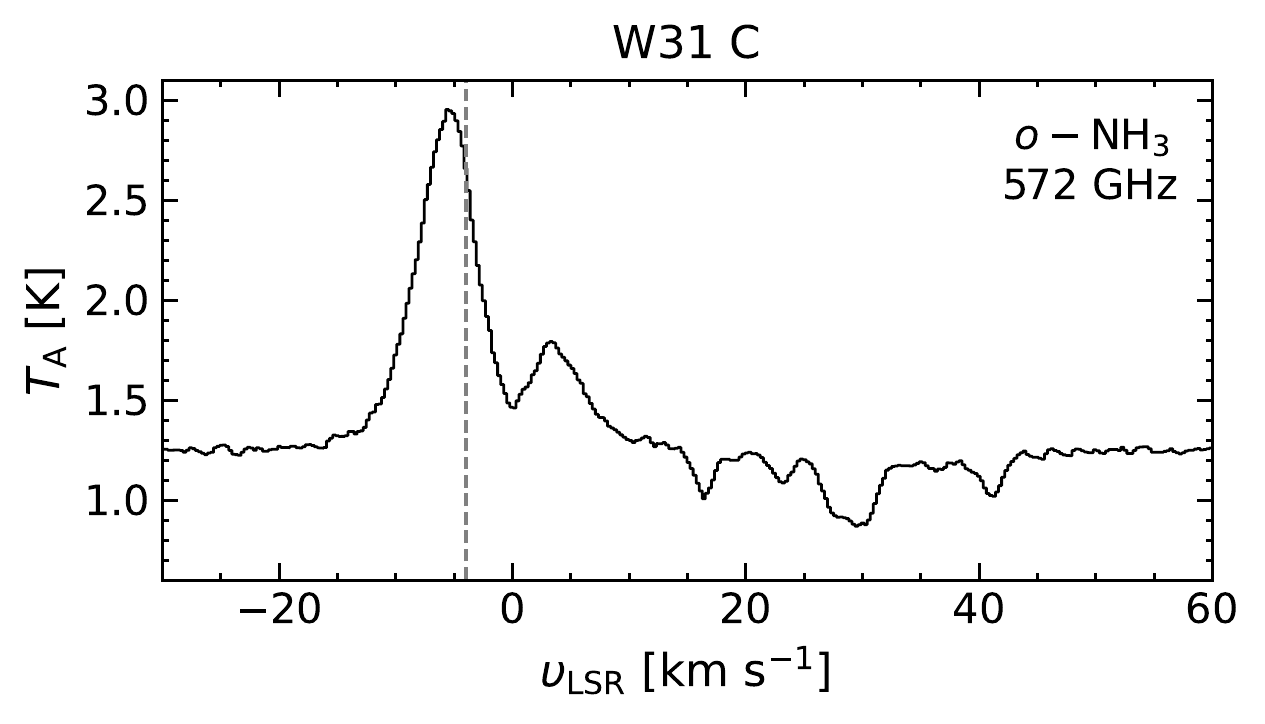}
    \includegraphics[width=0.425\textwidth]{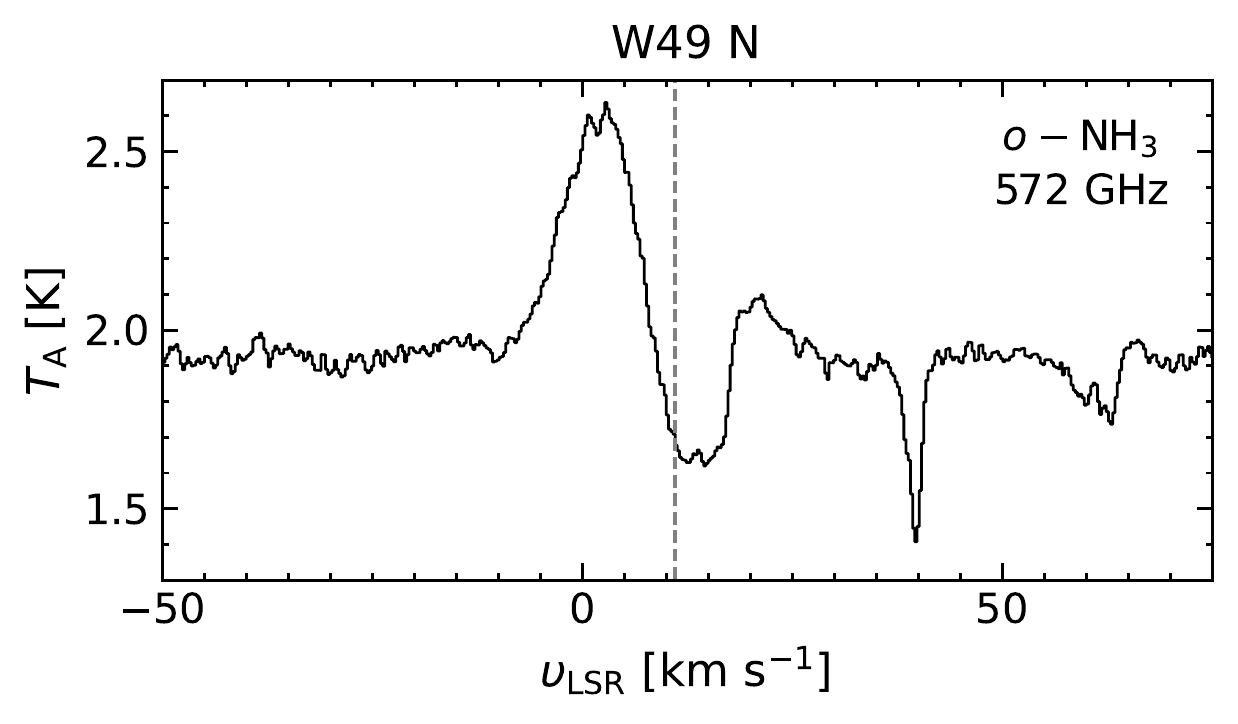}
    \caption{The ortho-NH$_3$(1$_0$--0$_0$) line towards W31~C (top) and W49~N (bottom) observed using \textit{Herschel}/HIFI. Vertical dashed grey lines mark the systemic velocities of the two sources.}
    \label{fig:NH3_W31C_W49N}
\end{figure}

\begin{figure}
    \centering
    \includegraphics[width=0.45\textwidth]{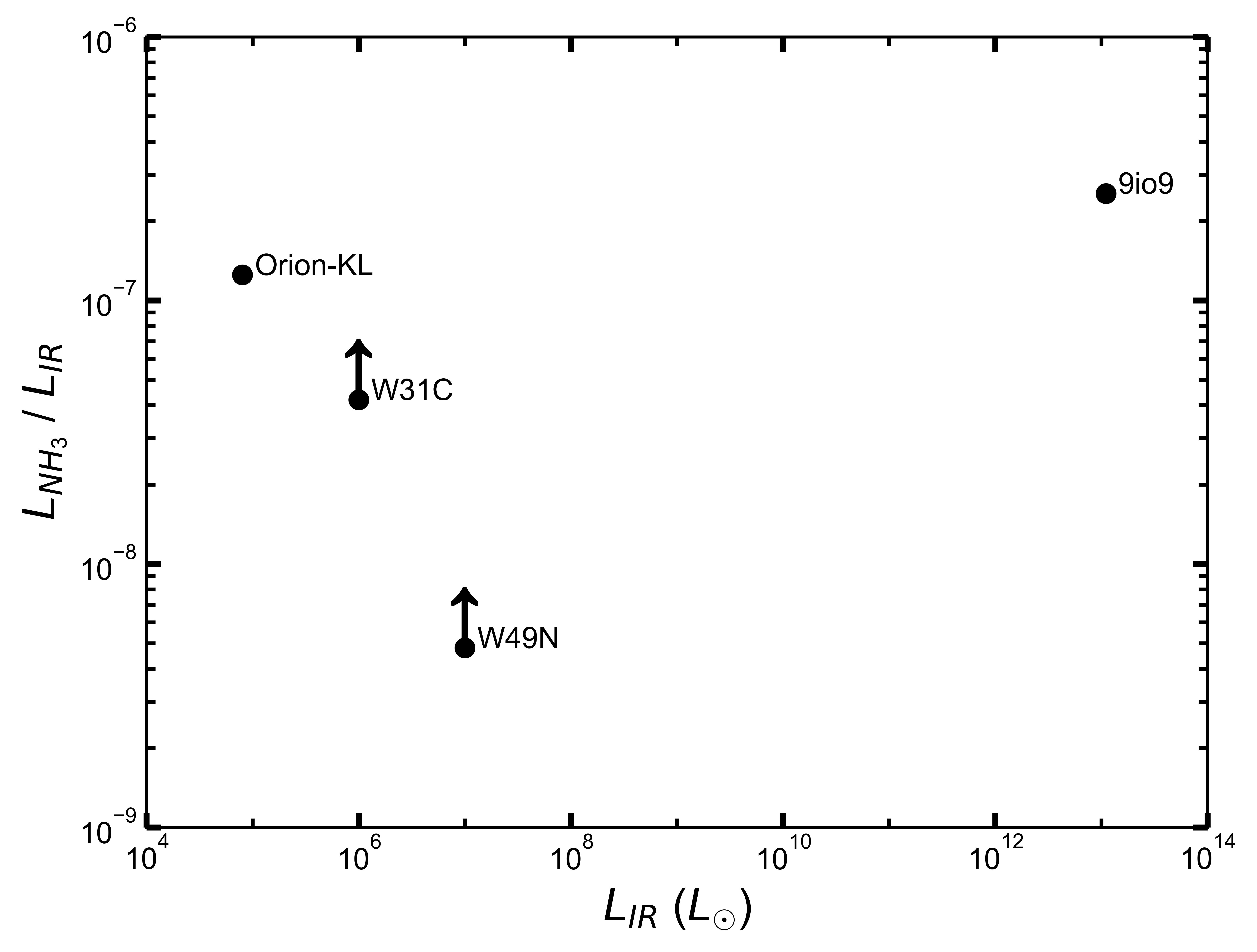}
    \caption{A comparison of the $L_{\rm NH_{3}}/L_{\rm IR}$ ratio versus $L_{\rm IR}$ for 9io9 at $z=2.6$ and a small sample of Galactic sources where \amm{} is detected. The {\it x}-axis spans ten orders of magnitude in infrared luminosity, whereas the luminosity ratios are consistent within a factor of 10.}
    \label{fig:LNH3-LIR}
\end{figure}

While the Orion molecular clouds have been extensively studied \citep[e.g.][]{GenzelStutzki1989}, we note that it has been argued that the energetics of Orion-KL are not dominated by high mass star formation, but rather by an  explosion \citep{Zapata2011}. Although one would expect a correspondingly high density of supernovae in 9io9, Orion-KL is arguably not a typical region in which presently \textit{high mass stars} are forming. In contrast W31\,C (G10.6$-$4) \citep[$L_{\rm IR}\sim\!10^{6}L_{\odot}$][]{} and the `mini-starburst' W49 N \citep[$L_{\rm IR}\sim\!10^{7}L_{\odot}$][]{Wright1977} are luminous Galactic high-mass star-forming regions located at distances of 4.8~kpc and 11.2~kpc, for which observations of the \amm{} line
have been published \citep{Persson2010, Persson2012}. Toward both sources, the spectra were taken as part of the \textit{Herschel} key guaranteed time project PRISMAS\footnote{PRobing InterStellar Molecules with Absorption line Studies (PI: M. Gerin)} using the Heterodyne Instrument for the Far-Infrared (HIFI). Unlike the corresponding para-NH$_3$ lines, which show almost exclusively absorption, toward both W49 N and W31 C, the spectra of the ortho-\amm{} line is far more complex, displaying strong emission at the velocities of the background sources with self-absorption features slightly offset from the systemic velocities (Figure~\ref{fig:NH3_W31C_W49N}). We model the emission by fitting Gaussian profiles centred at the systemic velocity of each source with the line widths optimised to fit the emission wings. Subsequently, the self-absorption features modelled using narrower Gaussian profiles centred at ${0}$~km~s$^{-1}$ and 13~km~s$^{-1}$ for W31 C and W49 N, respectively, are removed. The resulting fits are then used to derive integrated line intensities of 6.33~K~km~s$^{-1}$ and 1.33~K~km~s$^{-1}$ for W31 C and W49 N, respectively. Using a conversion factor of 482~Jy/K this yields line luminosities $L_{\rm NH_{3}}= 0.042L_{\odot}$ and $L_{\rm NH_{3}}=0.048L_{\odot}$, and $L_{\rm NH_{3}}/L_{\rm IR}$ of $4.2\times10^{-8}$ and $4.8\times10^{-8}$ toward W31 C and W49 N, respectively. The NH$_{3}$ line luminosities  should be considered lower limits due to uncertainties in the line intensities estimated and the nature of the observed self-absorption. If there is significant self-absorption of the NH$_{3}$ emission when averaged over galaxy scales in 9io9, then our measured luminosity could also be considered a lower limit. In Table~\ref{tab} and Figure~\ref{fig:LNH3-LIR} we compare the luminosity ratios we derive for the Galactic sources with 9io9. The ratios are broadly consistent within a factor of a few, despite the fact that the 9io9 measurement is galaxy-integrated across a system with an overall rate of star formation several orders of magnitude greater than the Milky Way.

\section{Conclusions}

The actual structure of star-forming regions in gas-dominated, high-redshift discs such as 9io9 remains unclear. While some studies have argued for the presence of `giant clumps' with properties similar to the cores of local Giant Molecular Clouds, but scaled up to sizes of order 100\,pc \citep[e.g.][]{rybak_alma_2015,hatsukade_high-resolution_2015}, others have pointed out that the reality of such features is questionable, and that star formation may well be smoother, or structured on smaller scales than can be reliably imaged interferometrically, even with the assistance of lensing \citep{ivison_giant_2020}. Regardless, it is evident that in order to drive globally elevated star formation, a large fraction of the cold ISM must be driven to high densities.

The introduction of supersonic turbulence is a key mechanism to achieve high gas density fractions \citep{geach_molecular_2012}, with the dispersion of the log-normal distribution describing the molecular gas density sensitive to the 1-dimensional average Mach number: $\mathcal{M}=\sigma_v/c_s$, with $\sigma_v$ the gas velocity dispersion and $c_s$ the speed of sound in the medium \citep{padoan_stellar_2002}. In the local Universe, mergers drive up $\mathcal{M}$ \citep[e.g.][]{Narayanan_2011}, and is the primary mechanism for ultraluminous emission in galaxies \citep{solomon_molecular_2005}. 9io9 -- like many other high redshift starbursts -- does not appear to be undergoing a major merger \citep[although see ][]{liu_massive_2022}, but VDIs \citep{dekel_formation_2009,inoue_non-linear_2016} are a viable alternative mechanism for locally driving up $\mathcal{M}$ resulting in pockets of high-density gas, and therefore star formation, across the gas-dominated disc. Confirming this in practice will require reliable high-resolution imaging (noting the caveat referenced above for interferometric data) that could map out the relative distribution of dense molecular gas compared to the bulk reservoir. It is important to note that minor mergers and interactions can catalyse VDIs \citep[e.g.][]{swinbank_interstellar_2011,Saha_2018}. 

That a large fraction of the molecular ISM in 9io9 resembles environments like Orion-KL and other Galactic environments, with the broad kinematics of NH$_3$ consistent with ordered disc rotation across the full range of molecular gas densities, we can picture an ensemble of millions of `Orion-KLs' embedded throughout the compact disc of 9io9; perhaps individually unremarkable, but en masse driving globally high star formation. This echos the evocative picture \cite{rybak_full_2020} present of another dusty, star-forming lensed galaxy, SDP.81 ($z\approx3$): they describe the system as  `full of Orions', based on the similarity of the ISM conditions on sub-kpc scales in SDP.81 compared to Orion. Our results appear to support this picture, and highlight the utility of the fainter, heavy rotor tracers in revealing the structure of the high-density ISM that is physically proximate with active star formation in young massive galaxies.

\section*{Acknowledgements}

We are grateful to the anonymous referee for their constructive comments. We also thank Matus Rybak for useful discussions. M.J.D.\ and J.E.G.\ acknowledge support from the Royal Society. S.D.\ is supported by an STFC
Rutherford Fellowship. This paper makes use of the following ALMA data: ADS/JAO.ALMA\#2019.1.01365.S. ALMA is a partnership of ESO (representing its
member states), NSF (USA) and NINS (Japan), together with NRC (Canada),
MOST and ASIAA (Taiwan), and KASI (Republic of Korea), in cooperation with
the Republic of Chile. The Joint ALMA Observatory is operated by ESO,
AUI/NRAO and NAOJ. Funded by the Deutsche Forschungsgemeinschaft (DFG, German Research
Foundation) under Germany's Excellence Strategy --- EXC-2094 ---
390783311. This research has made use of the University of
Hertfordshire high-performance computing facility
(\url{http://stri-cluster.herts.ac.uk}). 

\vspace{-11pt}

\section*{Data Availability}
Data will be shared on reasonable request to the corresponding author.

\vspace{-11pt}




\bibliographystyle{mnras}
\bibliography{references1.bib} 


\bsp	
\label{lastpage}
\end{document}